\documentclass[%
 reprint,
superscriptaddress,
preprintnumbers,
nofootinbib,
 amsmath,amssymb,
 aps,
 prl,
longbibliography,
twocolumn 
]{revtex4}  

\usepackage[pdftex]{graphicx}
\usepackage[utf8]{inputenc}
\usepackage{dcolumn}
\usepackage{bm}
\usepackage{amsmath}

\usepackage[colorlinks,allcolors=blue,linktocpage]{hyperref}
\usepackage{cleveref}
\crefname{equation}{Eq.}{Eqs.}%

\begin{document}
\title{4D dS vacua from AdS vacua of type IIB string theory and the AdS distance conjecture}
\author{Cao H. Nam}
\email{nam.caohoang@phenikaa-uni.edu.vn}  
\affiliation{Phenikaa Institute for Advanced Study and Faculty of Fundamental Sciences, Phenikaa University, Yen Nghia, Ha Dong, Hanoi 12116, Vietnam}
\date{\today}

\begin{abstract} 
In order for string theory to be made compatible with the low-energy observations of a positive cosmological constant, there have been attempts to construct dS vacua in string theory which are particularly difficult to realize. Instead of attempting to find de Sitter (dS) vacuum solutions, we point to a new way to make string theory consistent with low-energy dS cosmology. In this way, string theory lives in an anti-de Sitter (AdS) vacuum (which is simple to construct) that exists only in the high-energy regime; however, as going to the low-energy scales where the heavy string excitations and Kaluza-Klein modes are integrated out, we show that the effective picture of string theory in lower dimensions would exhibit a 4D dS vacuum without needing to add additional structures such as anti-D3 branes. Additionally, we point to evidence from bottom-up physics for the strong version of the AdS distance conjecture realized from AdS vacua in string theory. This evidence hence supports the sharpening of the AdS distance conjecture as one of the universal features of quantum gravity.
\end{abstract}

\maketitle

\emph{Motivations.}---String theory has been considered as one of the candidates for a correct theory of quantum gravity. However, it has been faced with the challenges of matching with the experimental observations at low-energy scales. One of them is to construct de Sitter (dS) vacua motivated by the observation of positive cosmological constant \cite{Weinberg2013}. However, it has been known that dS vacua in string theory are particularly difficult to realize. No-go theorems imply that there are no dS vacua in supergravity and string theory if the internal space is static, is compact, and has no singularities \cite{Gibbons2003,Maldacena2001}. The difficulties of the dS vacuum construction come also from explicitly computing string loop, higher-derivative, and nonperturbative corrections. The  Kachru-Kallosh-Linde-Trivedi construction allows anti-de Sitter (AdS) vacua uplifted to metastable dS vacua by adding anti-D3 branes \cite{Kachru2003,Burgess2003,Choi2005,Westphal2007}. But, there have been recent results in the literature which point to the problems with this dS vacuum construction regarding the backreaction of anti-D3 branes on the internal geometry \cite{Bena2010,Danielsson2015,Michel2015,Cohen-Maldonado2016,Gautason2016,Danielsson2017} and on the 4D moduli \cite{Moritz2018}, and nonsupersymmetric (non-SUSY) Giddings-Kachru-Polchinski solutions (derived in Ref. \cite{Giddings2002}) \cite{Sethi2018}. In addition, there have been some attempts at embedding dS cosmology within string theory \cite{Banerjee2018,Alwis2021}. The technical difficulties have hence suggested the possibility that string theory admits no dS vacua which do not suffer from instabilities at all \cite{Brennan2017}. Hence, the dS conjecture \cite{Obied2018} was proposed as well as the dS instability \cite{Garga2019} was studied to realize no dS vacua in string theory. For reviews about the status of the dS vacuum construction in string theory, see \cite{Danielsson2018,Cicoli2019}.

On the contrary, AdS vacua in string theory are understood very well and simple to construct. And, another fact is that the dS vacuum that we observe has been realized in the low-energy regime so far. This means that it is not known whether the vacuum is still dS in the high-energy regime; in other words, it is possible that the vacuum would be AdS at the high-energy scales. These facts imply an ideal that makes string theory compatible with the low-energy observations of positive vacuum energy without needing to find its dS vacua as follows. We start from the well-known AdS vacua of type IIB string theory with the compactification geometry given by $M_5\times X_5$ where solving the stringy 10D equations of motion would lead to $M_5$ to be an $\text{AdS}_5$ factor, which exists only in the high-energy regime. In addition, we consider the compactification of $M_5$ on a circle in order to obtain the observed 4D world where the 4D tensor component of the $M_5$ metric is in general dependent on the fifth coordinate. We point to that the non-trivial dynamics of the 4D tensor component of the $M_5$ metric along the fifth dimension leads to a positive contribution to energy in the 4D effective theory. As a result, a 4D dS vacuum can emerge in the low-energy regime from an AdS vacuum of the higher-dimensional theory existing at the high-energy scales.

A feature of AdS vacua in string theory is that an infinite tower of states becomes light in the limit of the AdS curvature radius going to infinity due to no scale separation between the AdS curvature radius and the radius of the internal space. This implies the proposal of the AdS distance conjecture \cite{Lust2019} which is a generalization of the swampland distance conjecture \cite{Ooguri2007} to the metric configuration space and is stated as follows: The near-flat limit of any AdS vacuum is accompanied by an infinite tower of states whose mass scale behaves as $m\sim|\Lambda|^\alpha$ with $\Lambda$ to be a cosmological constant and $\alpha\geq\frac{1}{2}$ required by the strong version. This conjecture together with other swampland conjectures have been used to place the constraints on the effective theories which can be completed consistent with quantum gravity in the ultraviolet \cite{Vafa2005,Palti2019,Grana2021}, and interesting implications for the neutrino masses, cosmological constant, and electroweak vacuum have been found \cite{Martin-Lozano2017,Valenzuela2017,Hamada2017,Gonzalo2018a,Gonzalo2018b,Rudelius2021,Gonzalo2021,Nam2022b}. However, evidence coming from the bottom-up physics for the strong version of the AdS distance conjecture, which is necessary to test this conjecture and sharpens it (and thus the swampland distance conjecture) as one of the universal features of quantum gravity, is still missing. We will show that the mass spectrum of the KK tower for the 5D bulk fields represents bottom-up evidence for the strong version of the AdS distance conjecture. This result hence supports that the AdS distance conjecture (and hence the swampland distance conjecture) realized from the AdS vacuum construction in string theory can be applied in general for quantum gravity.

\emph{AdS vacua in type IIB string theory.}---Our starting point is the 10D action for the massless string excitations of type IIB string theory, which is given by
\begin{eqnarray}
S=\frac{1}{\kappa^2_{10}}\int d^{10}X\sqrt{-g}\left[\mathcal{R}-\frac{1}{2}\left(\nabla\Phi\right)^2-\frac{g^2_s}{2}e^{a_p\Phi}F^2_{p+2}\right],\label{10Daction}
\end{eqnarray}
where $\kappa^2_{10}\equiv(2\pi)^7g^2_sl^8_s$ with $g_s$ and $l_s$ to be the string coupling and the string length, respectively, the dilaton coupling parameter $a_p$ for the Ramond-Ramond sector is $a_p=(3-p)/2$, and $F^2_{p+2}\equiv F_{M_1M_2...M_{p+2}}F^{M_1M_2...M_{p+2}}/(p+2)!$ with $F_{M_1M_2...M_{p+2}}$ to be the field strength tensor of $(p+1)$-form gauge fields. We consider the solution of (\ref{10Daction}) with the following geometry
\begin{eqnarray}
ds^2_{10}&=&ds^2_{M_5}+L^2ds^2_{X_5},\nonumber\\
e^\Phi&=&g_s,\label{IIBgeometry}\\
g_sF_5&=&\alpha(1+\star)\text{vol}_{X_5},\nonumber
\end{eqnarray}
where $X_5$ is a 5D Sasaki-Einstein manifold \cite{Gubser1999} which has the curvature satisfying $\mathcal{R}_{X_5}=20$ and is stabilized by $N$ units of flux, $\text{vol}_{X_5}$ refers to the five-form volume of $X_5$, $\alpha=16\pi g_sNl^4_s(\pi^3/\text{Vol}_{M_5})$ which is determined by the flux quantization constraint $\int_{X_5}\star F_5=(2\pi l_s)^{-3}N\kappa^2_{10}/(g_sl_s)$ with $\text{Vol}_{X_5}$ to be the volume of $X_5$, and other $(p+1)$-form fields are trivial. 

The equations of motion $\square\Phi=g^2_sa_pe^{a_p\Phi}F^2_{p+2}/[2(p+2)!]$ and $\triangledown_M\left(e^{a_p\Phi}F^{MN_1...N_{p+1}}\right)=0$ are satisfied for the constant dilaton and self-dual five-form field [and other $(p+1)$-form fields which vanish], respectively. Whereas, Einstein field equations ${\mathcal{R}_M}^N=F_{MM_1M_2M_3M_4}F^{NM_1M_2M_3M_4}/96$ lead to 
\begin{eqnarray}
L^4&=&4\pi g_sNl^4_s\frac{\pi^3}{\text{Vol}_{X_5}},\label{AdSrad}\\
\mathcal{R}_{M_5}&=&-\frac{20}{L^2},\label{5Deom}
\end{eqnarray}
where $\mathcal{R}_{M_5}$ denotes the scalar curvature of $M_5$. Eqs. (\ref{AdSrad}) and (\ref{5Deom}) together with the geometry (\ref{IIBgeometry}) imply that an exact background of type IIB string theory which is obtained from solving the stringy 10D equations of motion is a factor $\text{AdS}_5$ times a 5D internal manifold, i.e., $\text{AdS}_5\times X_5$.\footnote{The well-known case is that $X_5$ is a five-sphere $S^5$ related to AdS/CFT correspondence \cite{Maldacena1998}.}

\emph{Dimensionally reduced action of string theory.}---With the background geometry (\ref{IIBgeometry}), reducing the 10D action (\ref{10Daction}) of type IIB string theory on $X_5$ we get the following 5D effective action as follows
\begin{eqnarray}
S_{5D}&=&\frac{M^3_5}{2}\int d^5X\sqrt{-g_5}\left[\mathcal{R}_{M_5}-2\Lambda\right],\label{5Dgraact}
\end{eqnarray}
where $M^3_5=2\text{Vol}_{M_5}L^5/\kappa^2_{10}$ and $\Lambda=-6/L^2$ with $L$ as given by (\ref{AdSrad}). This action means that the 5D effective theory of string theory would be in the AdS vacuum. 

In addition, in order to obtain the 4D observed world we consider the compactification of $M_5$ on a circle $S^1$ where the most general setting of this compactification is given by a principal bundle with the typical fiber to be $U(1)$ \cite{Coquereaux1988,Bailin1987,Overduin1997}, which adopts the local coordinates as $\left(x^\mu,\theta\right)$ with $\{x^\mu\}\in\mathbb{R}^4$ and $\theta$ being an angle parametrizing the fifth dimension of $M_5$ corresponding to the coordinate transformation as $x^\mu\rightarrow x'^\mu=x'^\mu(x)$ and $\theta\rightarrow\theta'=\theta+\alpha(x)$. Hence, the metric equipped on $M_5$ takes the following general form
\begin{eqnarray}
ds^2_{M_5}=g_{\mu\nu}dx^\mu dx^\nu+R^2\left[d\theta+g_{_A}A_\mu dx^\mu\right]^2,\label{KKmetric}
\end{eqnarray}
where $g_{\mu\nu}$, $A_\mu$, and $R$ are the 4D tensor, 4D vector, and 4D scalar component fields of the bulk metric on $M_5$, respectively, and $g_{_A}$ is the corresponding gauge coupling. With this ansatz, we can explicitly expand $\mathcal{R}_{M_5}$ given in the action (\ref{5Dgraact}) in terms of the 4D component fields (see Appendix A for the detailed computation) as follows
\begin{eqnarray}
\mathcal{R}_{M_5}&=&\hat{\mathcal{R}}+\frac{1}{4R^2}\left(\partial_\theta g^{\mu\nu}\partial_\theta g_{\mu\nu}+g^{\mu\nu}g^{\rho\lambda}\partial_\theta g_{\mu\nu}\partial_\theta g_{\rho\lambda}\right)\nonumber\\
&&-\frac{g^2_{_A}R^2}{4}F_{\mu\nu}F^{\mu\nu},\label{curexp}
\end{eqnarray}
where $\hat{\mathcal{R}}\equiv g^{\mu\nu}(\hat{\partial}_\lambda\Gamma^\lambda_{\nu\mu}-\hat{\partial}_\nu\Gamma^\lambda_{\lambda\mu}+\Gamma^\rho_{\nu\mu}\Gamma^\lambda_{\lambda\rho}-\Gamma^\rho_{\lambda\mu}\Gamma^\lambda_{\nu\rho})$ with $\Gamma^\rho_{\mu\nu}\equiv\frac{g^{\rho\lambda}}{2}(\hat{\partial}_\mu g_{\lambda\nu}+\hat{\partial}_\nu g_{\lambda\mu}-\hat{\partial}_\lambda g_{\mu\nu})$, $\hat{\partial}_\mu\equiv\partial_\mu-g_{_A}A_\mu\partial_\theta$, and $F_{\mu\nu}=\partial_\mu A_\nu-\partial_\nu A_\mu$.

\emph{The emergence of dS vacuum in 4D world.}---In the previous section, we have seen that type IIB string theory exists at the AdS vacuum. However, in the following we will show that this AdS vacuum exhibits only in the high-energy regime. But, when approaching the low-energy regime, the effective theory of type IIB string theory in lower dimensions would exhibit a 4D dS vacuum which is consistent with the low-energy observations. The emergence of dS vacuum here is essentially due to the presence of the second term in Eq. (\ref{curexp}) which has been ignored in the literature, because the $\theta$ dependence of the 4D tensor component of the bulk metric equipped on $M_5$ is usually not considered.

Let us first obtain the bulk profile of the $4$D tensor component which describes its dynamics along the fifth dimension of $M_5$. By varying the action (\ref{5Dgraact}) in the 4D tensor component of the bulk metric, we find the following equation
\begin{eqnarray}
\bar{\mathcal{R}}_{\mu\nu}-\frac{1}{2}g_{\mu\nu}\bar{\mathcal{R}}+\Lambda g_{\mu\nu}+\frac{1}{4R^2}\left[g_{\mu\rho}g_{\nu\lambda}\partial^2_\theta g^{\rho\lambda}\right.&&\nonumber\\
\left.-\partial^2_\theta g_{\mu\nu}+g^{\rho\lambda}\partial_\theta g_{\rho\lambda}\partial_\theta g_{\mu\nu}+2\partial_\theta\left(g_{\mu\nu}g^{\rho\lambda}\partial_\theta g_{\rho\lambda}\right)\right.&&\nonumber\\
\left.-\frac{1}{2}g_{\mu\nu}\left\{\partial_\theta g^{\rho\lambda}\partial_\theta g_{\rho\lambda}-\left(g^{\rho\lambda}\partial_\theta g_{\rho\lambda}\right)^2\right\}\right]=0,&&\label{4Dtensor-equ}
\end{eqnarray}
where $\bar{\mathcal{R}}_{\mu\nu}\equiv(\partial_\lambda\bar{\Gamma}^\lambda_{\nu\mu}-\partial_\nu\bar{\Gamma}^\lambda_{\lambda\mu}+\bar{\Gamma}^\rho_{\nu\mu}\bar{\Gamma}^\lambda_{\lambda\rho}-\bar{\Gamma}^\rho_{\lambda\mu}\bar{\Gamma}^\lambda_{\nu\rho})$ with $\bar{\Gamma}^\rho_{\mu\nu}\equiv\frac{g^{\rho\lambda}}{2}(\partial_\mu g_{\lambda\nu}+\partial_\nu g_{\lambda\mu}-\partial_\lambda g_{\mu\nu})$ and $\bar{\mathcal{R}}\equiv g^{\mu\nu}\bar{\mathcal{R}}_{\mu\nu}$. We have here obtained Eq. (\ref{4Dtensor-equ}) in the vacuum $R=\text{const}$ and $A_\mu=0$ which are the solution of their equations of motion as seen late. We separate the variables as $g_{\mu\nu}(x,\theta)=\chi(\theta)g^{(4)}_{\mu\nu}(x)$ where $g^{(4)}_{\mu\nu}(x)$ is identified as the usual metric in the 4D effective theory and $\chi(\theta)$ is its profile. Then, we find
\begin{eqnarray}
\mathcal{R}^{(4)}_{\mu\nu}-\frac{1}{2}g^{(4)}_{\mu\nu}\mathcal{R}^{(4)}+\Lambda_4g^{(4)}_{\mu\nu}&=&0,\label{effEinsEq}\\
3\chi''(\theta)+8\frac{\chi'(\theta)^2}{\chi(\theta)}+\frac{2\Lambda}{R^{-2}}\chi(\theta)&=&\frac{2\Lambda_4}{R^{-2}},\label{chiEq}
\end{eqnarray}
where $\mathcal{R}^{(4)}_{\mu\nu}$ and $\mathcal{R}^{(4)}$ are the usual Ricci and scalar curvatures of the 4D effective geometry of spacetime written in terms of $g^{(4)}_{\mu\nu}(x)$, respectively, and $\Lambda_4$ is a constant. It is important to remark that Eq. (\ref{chiEq}) is a nonlinear differential equation and hence the solution of $g_{\mu\nu}(x,\theta)$ should not be given as the linear combination of partial solutions. Whereas, Eq. (\ref{effEinsEq}) determines the 4D effective geometry of spacetime sourced by a cosmological constant $\Lambda_4$ which is originated from the dynamics of the 4D tensor component of the bulk metric along the fifth dimension of $M_5$. 

To find an analytical solution for Eq. (\ref{chiEq}) with the boundary condition $\chi(-\pi)=\chi(\pi)$ \cite{boundcon} for the general value of $\Lambda_4$ is a difficult task. However, with $\Lambda_4=0$ corresponding to the situation of small $\Lambda_4$, it is easy to find an analytical solution as follows
\begin{eqnarray}
\chi(\theta)=\cosh^{\frac{3}{11}}\left(\frac{\sqrt{22}}{3}\kappa\theta\right),\label{chisol}
\end{eqnarray}
where $\kappa\equiv\sqrt{-\Lambda/R^{-2}}$. For the $\Lambda_4\neq0$, a particularly analytical solution is found as
\begin{eqnarray}
\chi(\theta)=\frac{11\Lambda_4}{19|\Lambda|}\left[\cosh\left(\sqrt{\frac{2}{11}}\kappa\theta\right)-1\right].\label{chisol2}
\end{eqnarray}
Note that, due to the topology of $S^1$ the 4D metric component and thus its bulk profile $\chi(\theta)$ must be periodic with the period $2\pi$, i.e., $\chi(\theta)=\chi(\theta+2\pi)$. One can make the solution $\chi(\theta)$ periodic with the period $2\pi$ by reflecting it at the boundary as discussed in Appendix B.\footnote{In the case of $\Lambda>0$, the solution for $\chi(\theta)$ is related to the cosine function which is periodic and thus itself is compatible with the topology of $S^1$ \cite{Nam2022b,Nam2021}.}

The fact that $\chi(\theta)$ is non-negative implies $\Lambda_4\geq0$, which means that the non-trivial profile given by Eqs. (\ref{chisol}) and (\ref{chisol2}) for the 4D tensor component of the bulk metric along the fifth dimension of $M_5$ should give a non-negative contribution to the energy in the 4D effective theory as seen in Eq. (\ref{effEinsEq}). In this sense, the 4D effective theory of type IIB string theory would be at the dS vacuum.

A key question that here arises is what would lead to the non-trivial profile for the 4D tensor component? We can see that the essential point which leads to this non-trivial profile is due to the presence of the second term on the left-hand side of Eq. (\ref{chiEq}). This term comes from the non-linear property of the gravitational field: Gravity is itself a source that creates gravity. If this term is absent then Eq. (\ref{effEinsEq}) becomes linear and hence the solution would be a sum of all possible modes with the different values of $\Lambda_4$. However, the excitation modes with $\Lambda_4>0$ would decay to the lower modes with $\Lambda_4<0$. As a result, the 4D effective theory would exist at the negative energy state or the AdS vacuum. Therefore, we can realize that the non-trivial profile for the 4D tensor component leading to the emergence of the dS vacuum in the 4D effective theory is essentially due to its non-linear property.

\begin{figure}[t]
 \centering
\begin{tabular}{cc}
\includegraphics[width=0.4 \textwidth]{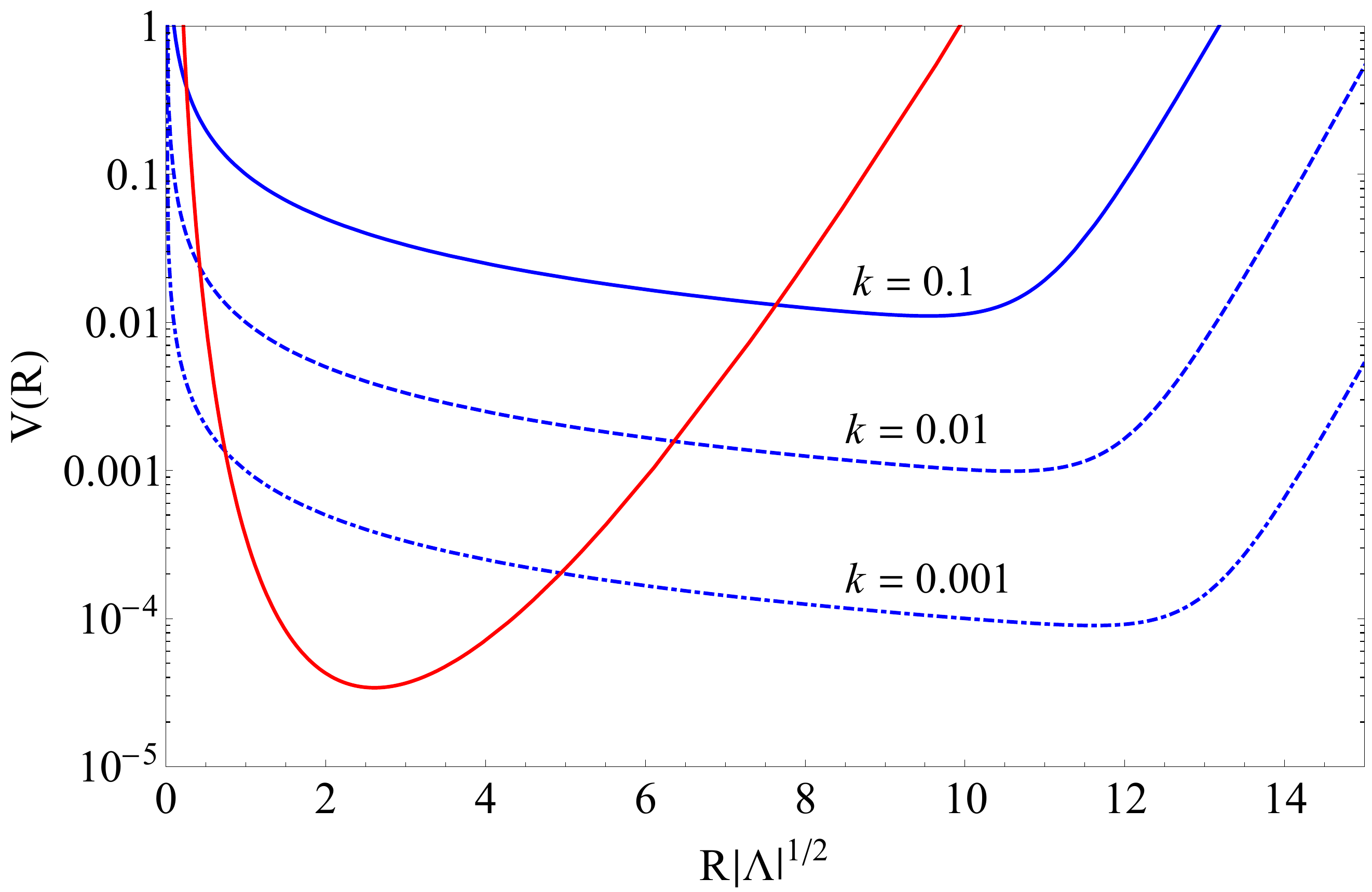}
\end{tabular}
 \caption{The radion potential for two cases: The red and blue curves correspond to the solutions (\ref{chisol}) and (\ref{chisol2}), respectively. Here, the radion potential corresponding to the red curve is rescaled by $|\Lambda|^3R^2_0$, $k\equiv|\Lambda|^{1/2}\Lambda_4R_0$, and we have considered $\sum_fn_f-\sum_bn_b=2$ and $m_{f,b}=0$ as a benchmark case.}\label{radion-pot}
\end{figure}

In order to show the 4D dS vacuum actually emerged in the effective picture of string theory in lower dimensions, we need to demonstrate that the ansatz (\ref{KKmetric}) with the vacuum configuration $g_{\mu\nu}=\chi(\theta)g^{(4)}_{\mu\nu}(x)$ (with the 4D metric $g^{(4)}_{\mu\nu}$ corresponding to the 4D dS geometry), $R=\text{const}$, and $A_\mu=0$ which has just been found above satisfies the 4D stringy equations of motion. The ansatz (\ref{KKmetric}) and Eq. (\ref{effEinsEq}) suggest that the dimensional reduction of the 5D action (\ref{5Dgraact}) on $S^1$ leads to the 4D effective action given in Einstein frame as follows
\begin{eqnarray}
S_{4D}&=&\int d^4x\sqrt{|g_4|}\left[\frac{M^2_{\text{Pl}}}{2}\left\{\mathcal{R}^{(4)}-\frac{3}{2}\left(\frac{\partial_\mu R}{R}\right)^2\right\}-V(R)\right.\nonumber\\
&&\left.-\frac{g^2_{_A}\pi M^3_5R^3}{4}F_{\mu\nu}F^{\mu\nu}-\frac{m^2_A}{2}A^\mu A_\mu\right],\label{4Deffact}
\end{eqnarray}
where $M^2_{\text{Pl}}=M^3_5R_0\int^\pi_{-\pi}d\theta\chi$, $m^2_A=3M^3_5R_0g^2_{_A}\int^\pi_{-\pi}d\theta[\chi''-\chi'^2/(2\chi)]$, and $V(R)$ refers to the potential of the radion field $R$. (In order to change to Einstein frame, we have rescaled $g^{(4)}_{\mu\nu}\rightarrow\Omega^{-2}g^{(4)}_{\mu\nu}$ with $\Omega^2=R/R_0$ and $R_0$ to be an arbitrary scale.) The radion potential $V(R)$ is given as follows
\begin{eqnarray}
V(R)=V_{\text{tr}}(R)+V_{\text{1L}}(R),
\end{eqnarray}
where $V_{\text{tr}}(R)$ is the tree-level term which is generated by the dynamics of $4$D tensor component along the fifth dimension of $M_5$ and is given by
\begin{eqnarray}
V_{\text{tr}}(R)=M^2_{\text{Pl}}\frac{R_0}{R}\Lambda_4,\label{clas-radpot}
\end{eqnarray}
and $V_{\text{1L}}(R)$ is the quantum level term which is generated by the (one loop) Casimir energy contribution and reads
\begin{eqnarray}
V_{\text{1L}}(R)=\sum_i(-1)^{s_i}n_iR\left(\frac{R_0}{R}\right)^2\rho_i(R)\int^\pi_{-\pi}d\theta\chi^2(\theta),
\end{eqnarray}
where $s_i=0(1)$ for the fermions(bosons), $n_i$ refers to the number of degrees of freedom with respect to the $i$-th particle, and the Casimir energy density is given by \cite{Arkani-Hamed2007}
\begin{eqnarray}
\rho_i(R)=\sum^{\infty}_{n=1}\frac{2m^5_i}{(2\pi)^{5/2}}\frac{K_{5/2}(2\pi nm_iR)}{(2\pi nm_iR)^{5/2}},
\end{eqnarray}
with $m_i$ and $K_{5/2}(z)$ to be the mass of the $i$-th particle and the modified Bessel function, respectively.\footnote{Here, the matter fields such as the Standard Model which are not contained in the field content of $\text{AdS}_5$ supergravity can be embedded by adding probe D7-branes wrapping around an internal cycle which is a submanifold of $M_5$ \cite{Karch2002,Gherghetta2006}} We can easily see that the radion potential behaves as $(\sum_fn_f-\sum_bn_b)/R^6$ and $(\sum_fn_f-\sum_bn_b)|_{m=0}\int^\pi_{-\pi}d\theta\chi^2(\theta)/R^6$ in the regions of $R\rightarrow0$ and $R\rightarrow\infty$, respectively, where $\sum_fn_f-\sum_bn_b$ ($(\sum_fn_f-\sum_bn_b)|_{m=0}$) is the net number of (massless) fermionic and bosonic degrees of freedom. This means that for $\sum_fn_f-\sum_bn_b>0$ and $(\sum_fn_f-\sum_bn_b)|_{m=0}>0$ the radion potential would approach the positive infinity for both $R\rightarrow0$ and $R\rightarrow\infty$ and hence there is always a stable minimum. Additionally, with the proper parameters this minimum has the positive energy corresponding to the dS vacuum, as depicted in Fig. \ref{radion-pot}.
The radion potential $V(R)$ thus allows us to fix physically the size of the fifth dimension of $M_5$ or the vacuum expectation value of the radion field $R$.

The 4D stringy equations of motion associated with the 4D effective action (\ref{4Deffact}) are derived in Appendix C. It is easy to see that the equations of motion for the radion field $R$ and the graviphoton $A_\mu$ lead to the vacuum solution $R=\text{const}$ and $A_\mu=0$ where the constant corresponds to the stable minimum of the radion potential $V(R)$. Whereas, the equations of motion for the 4D metric $g^{(4)}_{\mu\nu}$ is $\mathcal{R}^{(4)}_{\mu\nu}=\Lambda'_4g^{(4)}_{\mu\nu}$ where $\Lambda'_4\equiv M^{-2}_{\text{Pl}}V_{\text{sm}}$ with $V_{\text{sm}}$ referring to the stable minimum of the radion potential $V(R)$. Because $V_{\text{sm}}$ is positive with the proper parameters of the radion potential, the vacuum geometry of 4D effective theory is dS but the vacuum value is now shifted compared to Eq. (\ref{effEinsEq}) due to the radion stabilization.

Because the present construction is done relying on the 5D effective action obtained from the dimensional reduction of string theory on $X_5$, it is necessary to have a separation of the scales between the radius of $S^1$ and the scale $L$ of $X_5$. On the other hand, because of Eqs. (\ref{AdSrad}) and(\ref{5Deom}) the radius of $S^1$ must be much larger than the curvature radius of $\text{AdS}_5$, i.e. $R|\Lambda|^{1/2}\gg1$. We observe that the solution (\ref{chisol2}) can lead to the minimum of the radion potential fixing physically the radius of $S^1$ to satisfy this separation of the scales. Indeed, as seen in Fig. \ref{radion-pot} the minimum of the radion potential in the $R|\Lambda|^{1/2}$ direction would get larger when $k\equiv|\Lambda|^{1/2}\Lambda_4R_0$ decreases. This means that at a sufficiently small value of $k$ or $\Lambda_4$ it would yield the separation of the scales between the radius of $S^1$ and the scale of $X_5$.

Finally, let us emphasize that the transition from the AdS vacuum in the high-energy regime to the dS vacuum in the low-energy regime allows the effective theory to avoid the constraint of non-SUSY AdS conjecture \cite{Ooguri2017} because of the fact that supersymmetry is broken at the low-energy scales. If non-SUSY AdS vacuum is stable, then the effective theory would be in the swampland. On the contrary, non-SUSY AdS vacua would develop the non-perturbative instabilities and thus decay into SUSY vacua via the bubble nucleation \cite{Witten1981,Horowitz2007}.

\emph{AdS distance conjecture.}---We point to the first evidence from the bottom-up physics for the strong version of the AdS distance conjecture \cite{Lust2019}. In order to do this, we consider the 5D action (\ref{5Dgraact}) where the AdS radius $L$ is in general arbitrary instead of being given by Eq. (\ref{AdSrad}). This means that the 5D action (\ref{5Dgraact}) in this situation is not originated from the string theory compactification.

Let us first obtain the profile $Y_n(\theta)$ of the 5D bulk fields along the fifth dimension of $M_5$ corresponding to the solution (\ref{chisol}). The equations for the bulk profile of the 5D bulk fields are given in Appendix D. In general, it is not easy to obtain the analytical solutions to these equations. However, for small $\kappa$, we can find the analytical solutions to these equations by expanding $\chi(\theta)$ in $\kappa$. Up to the order $\kappa^2$, the solution form of these equations satisfying the boundary condition $Y(-\pi)=Y(\pi)$ \cite{boundcon} is found as
\begin{eqnarray}
Y_n(\theta)&=&\left[N_nH_{2n}\left(\sqrt{\frac{b_n}{3}}\theta\right)+{_1F_1}\left(-n;\frac{1}{2};\frac{b_n}{3}\theta^2\right)\right]\nonumber\\
&&\times\exp\left\{-a_n\theta^2\right\},\label{matt-prof}
\end{eqnarray}
where $n=0$, $1$, $2$, ..., $a_n$ and $b_n$ are parameters depending on $n$, $\kappa$, $R$, and the bulk mass, $H_{2n}(z)$ and ${_1F_1}(a;b;z)$ are the Hermite polynomial and the confluent hypergeometric function, respectively, and $N_n$ are the normalization constants. 

The degree of Hermite polynomial in the expression of $Y_n(\theta)$ must be even as a result of the boundary condition $Y_n(-\pi)=Y_n(\pi)$. From this, we can obtain the mass spectrum of the KK tower as follows
\begin{eqnarray}\label{KKspec}
m^2_n=\frac{|\Lambda|}{3}\times\left\{%
\begin{array}{ll}
    \left[\frac{5}{2}+4c_n+\frac{1+4n}{2}\sqrt{25+8c_n}\right], & \hbox{\text{scalar};} \\
    \left[1+4c_n+(1+4n)\sqrt{7+2c_n}\right], & \hbox{\text{fermion};} \\
    \left[\frac{3}{2}+4c_n+\frac{1+4n}{2}\sqrt{9+8c_n}\right], & \hbox{vector,} \\
\end{array}%
\right.\label{KKspectr}\nonumber\\
\end{eqnarray}
which correspond to the scalar, fermion, and vector fields, respectively, where $c_n\equiv n(1+2n)$. It should be noted here that first, we have considered the scalar field with zero bulk mass. Second, the mass of the bulk fermion must be zero due to the boundary condition on the bulk profile, which implies that the left- and right-handed components of the bulk fermion decouple and this is consistent with the fact that the compactification of spacetime on the circle $S^1$ breaks $SO(1,4)$ to $SO(1,3)\times U(1)$. Eq. (\ref{KKspec}) suggests that the mass of KK tower behaves in the power-law in the cosmological constant as $m_{KK}\sim|\Lambda|^{\alpha}$ with $\alpha=\frac{1}{2}$ which becomes light in the limit $\Lambda\rightarrow0$. This is consistent with the strong version of the AdS distance conjecture.

The above conclusion is still true in the general case. We observe from Fig. \ref{radion-pot} that the minimum of the radion potential $V(R)$ would fix $R|\Lambda|^{1/2}$. This implies $R^{-1}\propto|\Lambda|^{1/2}$ which means that the mass of the KK tower would be proportional to $|\Lambda|^{1/2}$. Hence, taking the cosmological constant $\Lambda$ to be zero would encounter an infinite tower of light states.

\emph{Conclusions.}---To achieve dS vacua in string theory motivated by the low-energy observations of a positive cosmological constant has proved to be a particularly difficult endeavor. Contrary to this, AdS vacua in string theory are common and simple to construct. Motivated by this fact and low-energy dS cosmology, we have indicated a new approach for the embedding of the observed dS vacuum into string theory which is one of the candidates for a unitary theory of quantum gravity. We do not attempt to find a dS vacuum solution in string theory, but we start from a well-known AdS vacuum with the compactification geometry given by $M_5\times X_5$. Then, we show that this AdS vacuum of string theory exists only in the high-energy regime. In other words, as approaching the low-energy scales where the heavy string excitations and KK modes are integrated out the effective picture of string theory in lower dimensions would exhibit a 4D dS vacuum (which has so far been observed in the low-energy regime) due to the non-trivial dynamics of the 4D tensor component of the $M_5$ metric along the fifth dimension which contributes positive energy in the 4D effective theory. This result clearly provides a new path in the construction of realistic models on AdS vacua (rather than dS vacua) in type IIB string theory, but it still leads to a dS vacuum in the 4D effective theory consistent with the low-energy observation of positive vacuum energy.

In addition, due to the non-trivial bulk profile of the 4D tensor component of the bulk metric, we revisited the mass spectrum of the KK tower for the 5D bulk fields. We showed that this mass spectrum satisfies the strong version of the AdS distance conjecture which is a generalization of the swampland distance conjecture and whose proposal was motivated by the realizations in the AdS vacuum construction in string theory. This offers evidence that comes from bottom-up physics and hence supports the AdS distance conjecture (and thus the swampland distance conjecture) as one of the universal features of quantum gravity.

\section*{Appendix A: The detail expansion of $\mathcal{R}_{M_5}$}

Eq. (\ref{curexp}) can be easily found if one works in the covariant frame $\Big\{\hat{\partial}_\mu,\partial_\theta\Big\}\equiv\left\{\partial_M\right\}$ where $\hat{\partial}_\mu$ and $\partial_\theta$ transform as a four-dimensional vector and a one-dimensional vector under the residual general coordinate transformation $x^\mu\rightarrow x'^\mu=x'^\mu(x)$ and $\theta\rightarrow\theta'=\theta+\alpha(x)$. In this frame, the coefficients of the Christoffel connection and the Riemann curvature tensor are given as follows
\begin{eqnarray}
\Gamma^P_{MN}&=&\frac{G^{PQ}}{2}\left(\partial_MG_{NQ}+\partial_NG_{MQ}-\partial_QG_{MN}\right)\nonumber\\
&&+\frac{G^{PQ}}{2}\left(C^O_{QM}G_{ON}+C^O_{QN}G_{OM}\right)+\frac{C^P_{MN}}{2},\nonumber\\
\mathcal{R}^O_{MPN}&=&\partial_P[\Gamma^O_{NM}]-\partial_N[\Gamma^O_{PM}]+
\Gamma^Q_{NM}\Gamma^O_{PQ}-\Gamma^Q_{PM}\Gamma^O_{NQ}\nonumber\\
&&-C^Q_{PN}\Gamma^O_{QM},
\end{eqnarray}
where $C^P_{MN}$ determines the commutation relation of any two frame fields as
\begin{equation}
\left[\partial_M,\partial_N\right]=C^P_{MN}\partial_P,
\end{equation}
and the metric and its dual are given in the coframe $\{dx^\mu,d\theta+g_{_A}A_\mu dx^\mu\}$ dual to $\{\hat{\partial}_\mu,\partial_\theta\}$ as
\begin{eqnarray}
  G_{MN}&=&\textrm{diag}\left(g_{\mu\nu},R^2\right),\nonumber\\
  G^{MN}&=&\textrm{diag}\left(g^{\mu\nu},R^{-2}\right).
\end{eqnarray}
We can explicitly expand the 5D scalar curvature $\mathcal{R}_{M_5}$ as follows
\begin{eqnarray}
\mathcal{R}_{M_5}=G^{MN}\mathcal{R}^P_{MPN}=g^{\mu\nu}\mathcal{R}^P_{\mu P\nu}+G^{\theta\theta}\mathcal{R}^P_{\theta P\theta},
\end{eqnarray}
where
\begin{eqnarray}
g^{\mu\nu}\mathcal{R}^P_{\mu P\nu}&=&g^{\mu\nu}\left(\partial_M\Gamma^M_{\nu\mu}-\hat{\partial}_\nu\Gamma^M_{M \mu}+\Gamma^N_{\nu\mu}\Gamma^M_{MN}-\Gamma^N_{M\mu}\Gamma^M_{\nu N}\right.\nonumber\\
&&\left.-C^N_{M\nu}\Gamma^M_{N\mu}\right)\nonumber\\
&=&g^{\mu\nu}\left(\hat{\partial}_\lambda\Gamma^\lambda_{\nu\mu}-\hat{\partial}_\nu\Gamma^\lambda_{\lambda\mu}+\Gamma^\rho_{\nu\mu}\Gamma^\lambda_{\lambda\rho}-\Gamma^\rho_{\lambda\mu}\Gamma^\lambda_{\nu\rho}\right)\nonumber\\
&&+g^{\mu\nu}\left(\partial_\theta\Gamma^\theta_{\nu\mu}+\Gamma^\theta_{\nu\mu}\Gamma^\lambda_{\lambda\theta}+\Gamma^\rho_{\nu\mu}\Gamma^\theta_{\theta\rho}+\Gamma^\theta_{\nu\mu}\Gamma^\theta_{\theta\theta}\right)\nonumber\\
&&-g^{\mu\nu}\left(\hat{\partial}_\nu\Gamma^\theta_{\theta\mu}+\Gamma^\rho_{\theta\mu}\Gamma^\theta_{\nu\rho}+\Gamma^\theta_{\lambda\mu}\Gamma^\lambda_{\nu\theta}+\Gamma^\theta_{\theta\mu}\Gamma^\theta_{\nu\theta}\right)\nonumber\\
&&-g^{\mu\nu}C^\theta_{\lambda\nu}\Gamma^\lambda_{\theta\mu},\label{sccurI}\\
G^{\theta\theta}\mathcal{R}^P_{\theta P\theta}&=&G^{\theta\theta}\left(\partial_M\Gamma^M_{\theta\theta}-\partial_\theta\Gamma^M_{M\theta}+\Gamma^N_{\theta\theta}\Gamma^M_{MN}-\Gamma^N_{M\theta}\Gamma^M_{\theta N}\right.\nonumber\\
&&\left.-C^N_{M\theta}\Gamma^M_{N\theta}\right)\nonumber\\
&=&G^{\theta\theta}\left(\hat{\partial}_\mu\Gamma^\mu_{\theta\theta}-\partial_\theta\Gamma^\mu_{\mu\theta}+\Gamma^\mu_{\theta\theta}\Gamma^\nu_{\nu\mu}+\Gamma^\theta_{\theta\theta}\Gamma^\mu_{\mu\theta}\right.\nonumber\\
&&\left.-\Gamma^\nu_{\mu\theta}\Gamma^\mu_{\theta\nu}-\Gamma^\mu_{\theta\theta}\Gamma^\theta_{\theta\mu}\right).\label{sccurII}      
\end{eqnarray}
In (\ref{sccurI}) and (\ref{sccurII}),  we find the following combinations
\begin{eqnarray}
g^{\mu\nu}\left(\hat{\partial}_\nu\Gamma^\theta_{\theta\mu}-\Gamma^\rho_{\nu\mu}\Gamma^\theta_{\theta\rho}\right)-G^{\theta\theta}\Gamma^\mu_{\theta\theta}\Gamma^\theta_{\theta\mu}&\equiv&\nabla_MY^M_1,\nonumber\\
g^{\mu\nu}\left(\partial_\theta\Gamma^\theta_{\nu\mu}+\Gamma^\theta_{\nu\mu}\Gamma^\lambda_{\lambda\theta}+\Gamma^\theta_{\nu\mu}\Gamma^\theta_{\theta\theta}\right)&\equiv&\nabla_MY^M_2\nonumber\\
&&+\frac{G^{\theta\theta}}{2}\partial_\theta g^{\mu\nu}\partial_\theta g_{\mu\nu},\nonumber\\
G^{\theta\theta}\left(\partial_\theta\Gamma^\mu_{\mu\theta}-\Gamma^\theta_{\theta\theta}\Gamma^\mu_{\mu\theta}\right)-g^{\mu\nu}\Gamma^\theta_{\mu\nu}\Gamma^\lambda_{\lambda\theta}&\equiv&\nabla_MY^M_3,\nonumber\\
G^{\theta\theta}\left(\hat{\partial}_\mu\Gamma^\mu_{\theta\theta}+\Gamma^\mu_{\theta\theta}\Gamma^\nu_{\nu\mu}+\Gamma^\mu_{\theta\theta}\Gamma^\theta_{\theta\mu}\right)&\equiv&\nabla_MY^M_4\nonumber\\
&&+\frac{g^{\mu\nu}}{2}\hat{\partial}_\mu G^{\theta\theta}\hat{\partial}_\nu G_{\theta\theta},\nonumber
\end{eqnarray}
where
\begin{eqnarray}
Y^M_1&\equiv&\left(g^{\mu\nu}\Gamma^\theta_{\theta\mu},0\right),\nonumber\\
Y^M_2&\equiv&\left(0,g^{\mu\nu}\Gamma^\theta_{\nu\mu}\right),\nonumber\\
Y^M_3&\equiv&\left(0,G^{\theta\theta}\Gamma^\mu_{\mu\theta}\right),\nonumber\\
Y^M_4&\equiv&\left(G^{\theta\theta}\Gamma^\mu_{\theta\theta},0\right).
\end{eqnarray}
Note that, $Y^M_1$ and $Y^M_4$ transform as a four-dimensional vector, whereas $Y^M_2$ and $Y^M_3$ transform as a one-dimensional vector. The terms $\nabla_MY^M_i$ with $i=1,2,3,4$ are divergences which correspond to the boundary terms and hence vanish at infinity.

\section*{\label{4DEOMs}Appendix B: The solution of $\chi(\theta)$ in different charts and its period}

\begin{figure}[t]
 \centering
\begin{tabular}{cc}
\includegraphics[width=0.4 \textwidth]{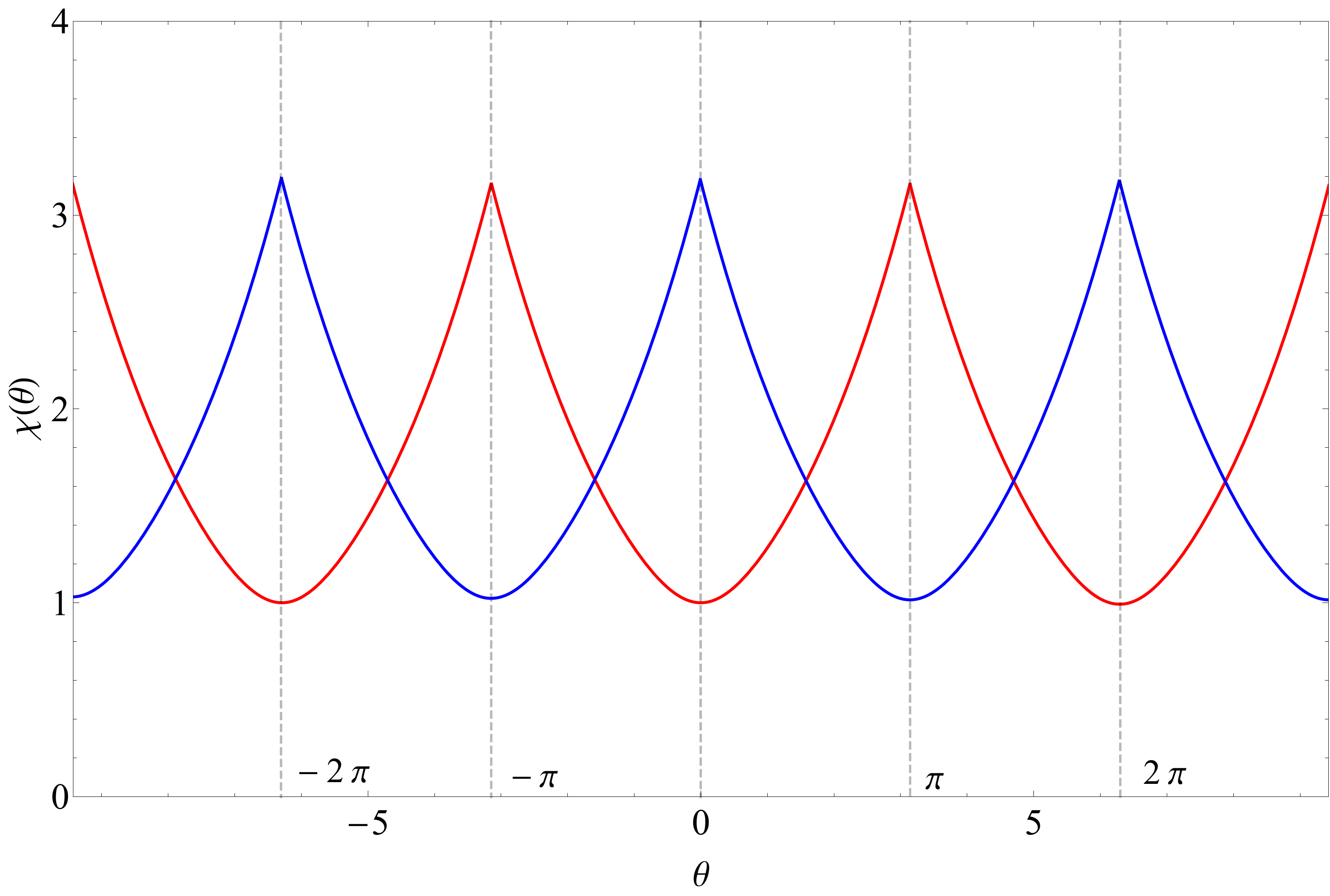}\\
\includegraphics[width=0.4 \textwidth]{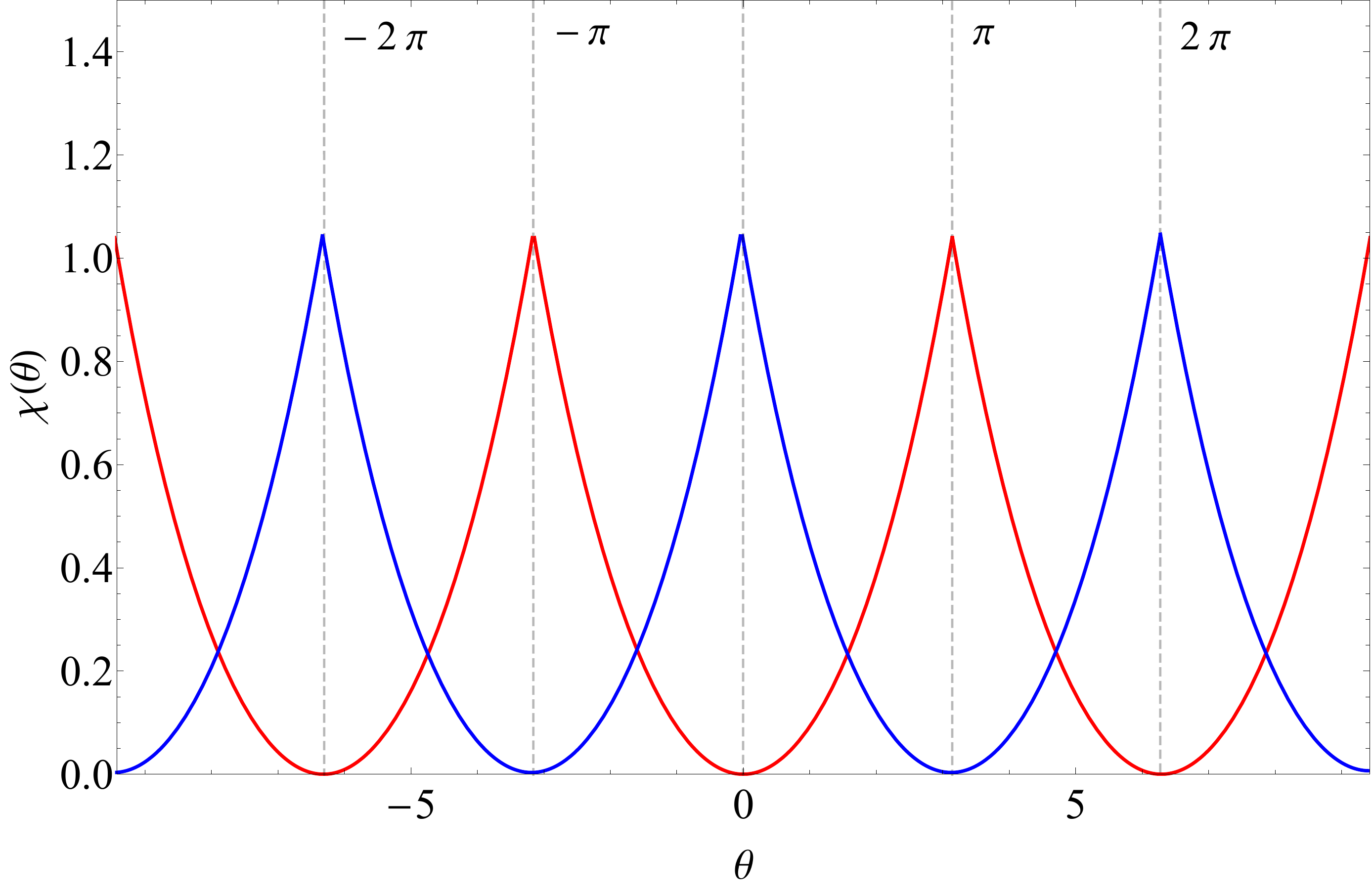}
\end{tabular}
 \caption{The solution of $\chi(\theta)$ in the charts $(-\pi,\pi)$ (red curve) and $(0,2\pi)$ (blue curve) which is periodic with the period $2\pi$ by reflecting it at the boundary. Top panel is for $\chi$ given in Eqs. (\ref{chisol1}) and (\ref{chisol3}). Bottom panel is for $\chi$ given in Eqs. (\ref{chisol2}) and (\ref{chisol4}).}\label{period-chi}
\end{figure}

In order to describe $S^1$, one needs at least two charts $(-\pi,\pi)$ and $(0,2\pi)$ which form an atlas of $S^1$. In chart $(-\pi,\pi)$, the solution of $\chi(\theta)$ satisfying the boundary condition $\chi(-\pi)=\chi(\pi)$ is given by 
\begin{eqnarray}
\chi(\theta)=\cosh^{\frac{3}{11}}\left(\frac{\sqrt{22}}{3}\kappa\theta\right),\label{chisol1}
\end{eqnarray}
for the case of $\Lambda_4=0$ and
\begin{eqnarray}
\chi(\theta)=\frac{11\Lambda_4}{19|\Lambda|}\left[\cosh\left(\sqrt{\frac{2}{11}}\kappa\theta\right)-1\right],\label{chisol2}
\end{eqnarray}
for the case of $\Lambda_4\neq0$. Whereas, in the chart $(0,2\pi)$ the solution of $\chi(\theta)$ satisfying the boundary condition $\chi(0)=\chi(2\pi)$ is given by 
\begin{eqnarray}
\chi(\theta)=\cosh^{\frac{3}{11}}\left[\frac{\sqrt{22}}{3}\kappa(\theta-\pi)\right],\label{chisol3}
\end{eqnarray}
for $\Lambda_4=0$ and
\begin{eqnarray}
\chi(\theta)=\frac{11\Lambda_4}{19|\Lambda|}\left[\cosh\left(\sqrt{\frac{2}{11}}\kappa(\theta-\pi)\right)-1\right],\label{chisol4}
\end{eqnarray}
for $\Lambda_4\neq0$. Because of the topology of $S^1$, the 4D metric component and thus its bulk profile $\chi(\theta)$ mus be periodic with the period $2\pi$, i.e. $\chi(\theta)=\chi(\theta+2\pi)$. We can make the solution $\chi(\theta)$ given above periodic with the period $2\pi$ by reflecting it at the boundary as depicted in Fig. \ref{period-chi}.
From this figure, we see that the bulk profile $\chi(\theta)$ is not differentiable at the boundary at $\theta=\pm\pi$ in the chart $(-\pi,\pi)$ but it is differentiable in the chart $(0,2\pi)$. Whereas, $\chi(\theta)$ is not differentiable at the boundary at $\theta=0,2\pi$ in the chart $(0,2\pi)$ but it is differentiable in the chart $(-\pi,\pi)$.

\section*{\label{4DEOMs}Appendix C: 4D stringy equations of motions}
In this appendix, we obtain the 4D stringy equations of motion from the variation of the action (\ref{4Deffact}) as
\begin{eqnarray}
G^{(4)}_{\mu\nu}&=&M^{-2}_{\text{Pl}}\left(T^A_{\mu\nu}+T^R_{\mu\nu}\right),\\
\nabla_\mu F^{\mu\nu}&=&-\frac{3\partial_\mu R}{R}F^{\mu\nu}+\frac{m^2_A}{\vartheta R^3}A^\nu\\
\square R&=&\frac{1}{R}(\partial_\mu R)^2+\frac{2R^2}{3M^2_{\text{Pl}}}\frac{\partial V}{\partial R}+\frac{\vartheta R^4}{2M^2_{\text{Pl}}}F_{\mu\nu}F^{\mu\nu}\nonumber\\
&&+\frac{R^2}{3M^2_{\text{Pl}}}\frac{\partial m^2_A}{\partial R}A_\mu A^\mu,
\end{eqnarray}
where $G^{(4)}_{\mu\nu}\equiv R^{(4)}_{\mu\nu}-\frac{1}{2}g^{(4)}_{\mu\nu}R^{(4)}$, $\vartheta\equiv g^2_{_A}\pi M^3_5$, and the energy-momentum tensors are given by
\begin{eqnarray}
T^A_{\mu\nu}&=&\vartheta R^3{F_\mu}^\lambda F_{\nu\lambda}+m^2_AA_\mu A_\nu-\frac{g^{(4)}_{\mu\nu}}{2}\left(\frac{\vartheta}{2} R^3F_{\sigma\lambda}F^{\sigma\lambda}\right.\nonumber\\
&&\left.+m^2_AA_\lambda A^\lambda\right),\\
T^R_{\mu\nu}&=&\frac{3M^2_{\text{Pl}}}{2R^2}\partial_\mu R\partial_\nu R-\frac{1}{2}g^{(4)}_{\mu\nu}\left[\frac{3}{2}\left(\frac{\partial_\lambda R}{R}\right)^2+\frac{2V(R)}{M^2_{\text{Pl}}}\right].\nonumber\\
\end{eqnarray}

\section*{Appendix D: Equations for the profile of the 5D bulk fields}

We here provide the equations determining the profile of the 5D bulk fields along the fifth dimension of $M_5$. For the 5D scalar field expanded as $\Phi(x,\theta)=\sum_n\phi^{(n)}(x)Y_n(\theta)$, the profile of the $n$th mode satisfies the following equation
\begin{eqnarray}
Y''_n+\frac{2\chi'}{\chi}Y'_n+R^2\left(\frac{m^2_n}{\chi}-m^2_\Phi\right)Y_n=0,\label{scal-prof}
\end{eqnarray}
where $m_\Phi$ and $m_n$ are the bulk mass and the mass of the $n$th mode, respectively. With respect to the 5D fermion expanded as
\begin{eqnarray}
\Psi(x,\theta)=\sum_n\left[\psi^{(n)}_{L}(x)Y_{Ln}(\theta)+\psi^{(n)}_{R}(x)Y_{Rn}(\theta)\right],
\end{eqnarray} 
the equations for the profile of the $n$th modes corresponding to left- and right-handed components are given as
\begin{eqnarray}
Y'_{Ln}+\left(\frac{m_\Psi}{R^{-1}}+\frac{\chi'}{\chi}\right)Y'_{Ln}-\frac{m_nR}{\chi^{1/2}}Y_{Rn}&=&0,\label{ferL-prof}\\
-Y'_{Rn}+\left(\frac{m_\Psi}{R^{-1}}-\frac{\chi'}{\chi}\right)Y'_{Rn}-\frac{m_nR}{\chi^{1/2}}Y_{Ln}&=&0,\label{ferR-prof}
\end{eqnarray}
where $m_\Psi$ and $m_n$ are the bulk mass and the Dirac mass of the $n$th mode. For the 5D gauge boson $A_M$ with the gauge $A_\theta=0$ and expanded as
\begin{eqnarray}
A_\mu(x,\theta)=\sum_nA^{(n)}_\mu(x)f_n(\theta),
\end{eqnarray}
an equation for the profile of the $n$th mode reads
\begin{eqnarray}
f''_{n}+\frac{\chi'}{\chi}f'_{n}+\frac{m^2_nR^2}{\chi}f_n=0.\label{gabos-prof}
\end{eqnarray}
The solution form of these equations are given in Eq. (18) where
\begin{eqnarray}
a_n&=&\frac{5\kappa^2}{12}\times\left\{%
\begin{array}{ll}
    \frac{4}{5}\left[1+\left(1+\frac{3m^2_nR^2}{4\kappa^2}\right)^{\frac{1}{2}}\right], & \hbox{\text{scalar};}\\
    \left[1+\left(1+\frac{12m^2_nR^2}{25\kappa^2}\right)^{\frac{1}{2}}\right], & \hbox{\text{fermion};}\\
    \frac{2}{5}\left[1+\left(1+\frac{3m^2_nR^2}{\kappa^2}\right)^{\frac{1}{2}}\right], & \hbox{\text{vector},}
\end{array}%
\right.
\end{eqnarray}
\begin{eqnarray}
b_n&=&\kappa^2\times\left\{%
\begin{array}{ll}
    2\left(1+\frac{3m^2_nR^2}{4\kappa^2}\right)^{1/2}, & \hbox{\text{scalar};}\\
    \frac{1}{2}\left(1+\frac{12m^2_nR^2}{25\kappa^2}\right)^{1/2}, & \hbox{\text{fermion};}\\
    \left(1+\frac{3m^2_nR^2}{\kappa^2}\right)^{1/2}, & \hbox{\text{vector},}
\end{array}%
\right.
\end{eqnarray}
with the mass $m_n$ corresponding to the $n$th mode for each of the field type given in Eq. (\ref{KKspectr}), and the normalization constants $N_n$ are computed from the orthonormal relation as
\begin{eqnarray}
\int^{\pi}_{-\pi}d\theta\chi(\theta)^{\omega}Y_n(\theta)Y_m(\theta)=\delta_{mn},
\end{eqnarray} 
with $\omega=1$, $3/2$, and $0$ for the scalar, fermion, and vector fields, respectively, and $Y_n(\theta)$ referring to the bulk profile of these fields.


\begin{thebibliography}{99}
\bibitem{boundcon} Because the point with $\theta=-\pi$ and $\theta=\pi$ is the same, any 5D field $\Psi(x,\theta)$ should satisfy the following boundary condition
\begin{eqnarray}
\Psi(x,-\pi)=\Psi(x,\pi).\nonumber
\end{eqnarray}
\end{thebibliography}
\end{document}